\documentclass[aps,prl,10pt,twocolumn,showpacs]{revtex4-1}
\usepackage{amsmath}
\usepackage{graphicx}
\usepackage{color}
\usepackage{braket}

\begin{document}
\title{Quasi-simultons in thermal atomic vapors}
\author{Thomas P.\ Ogden}
\author{K.\ A.\ Whittaker}
\author{J.\ Keaveney}
\author{S.\ A.\ Wrathmall}
\author{C.\ S.\ Adams}
\author{R.\ M.\ Potvliege}
\email{r.m.potvliege@durham.ac.uk}
\affiliation{Department of Physics, Joint Quantum Centre (JQC) Durham-Newcastle, Durham University, South Road, Durham DH1 3LE, United Kingdom}
\begin{abstract}
{The propagation of two-color laser fields through
optically thick atomic ensembles is studied. We demonstrate how the interaction between these two fields spawns the formation of
co-propagating, two-color
soliton-like pulses akin to the
simultons found by Konopnicki and Eberly 
[Phys.\ Rev.\ A {\bf 24}, 2567 (1981)]. For the particular case
of thermal Rb atoms, exposed to a combination
of a weak cw laser field resonant on the D1 transition
and a strong, sub-ns laser pulse resonant on the D2 transition,
simulton formation is initiated by an interplay between the 5s$_{1/2}$ -- 5p$_{1/2}$ and 5s$_{1/2}$ -- 5p$_{3/2}$ coherences which amplifies the D1 field at the arrival of the D2 pulse producing sech-squared pulse with a length of less than 10 microns.
This amplification is demonstrated in a time-resolved measurement of the light transmitted
through a thin thermal cell.
We find good agreement between experiment and a model that
includes the hyperfine structure of the relevant levels. With the addition of Rydberg dressing,
quasi-simultons offer interesting prospects for strong photon-photon interactions
in a robust environment.
} 
\end{abstract}
\maketitle

Self-induced transparency manifests dramatically by the formation of 
optical solitons propagating undistorted over
long distances in a medium opaque to a cw field of the same
wavelength \cite{McCall}.
A short light pulse may propagate as a soliton or split
into multiple solitons
only if it is sufficiently intense \cite{area}.
However, it has been known for some time that 
a weak soliton-like pulse at one frequency can co-propagate
with a stronger soliton-like pulse at another frequency.
Solutions of the Maxwell-Bloch equations describing this situation
in doubly resonant 3-level V-systems
were first found in the form of matched sech pulses,
or simultons,
under the condition that the two transitions
have the same oscillator strength \cite{Kono81,seealso}. Remarkably, the two pulses
may co-propagate as a simulton even if neither of them is strong enough to
support a soliton in the absence of the other field
\cite{Rahman}.
The condition that the two transitions have the same oscillator strength is difficult to realize experimentally \cite{othersystems} but
makes the Maxwell-Bloch equations
integrable in the sense of the inverse
scattering transform (in a suitable approximation),
which permits an in depth analysis of their
analytical solutions \cite{multisimultons}.

However, this condition
can be relaxed without compromising the formation of
pairs of soliton-like pulses co-propagating with little distortion over much longer distances
than allowed by Beer's law \cite{Bolshov85,Bolshov88}. We refer to such
pairs of pulses as quasi-simultons, to distinguish them from 
the ideal sech-simultons of Ref.~\cite{Kono81}.
It has been noted, in particular, that
a soliton on one transition of a V-system may enhance transmission of a weak pulse on the other
transition even if the latter
has a different oscillator strength \cite{Kozlov98}.
The soliton may amplify the weak pulse and transport it
through the medium
simulton-like \cite{KozlovandKozlova2009,KK2010}.
The term 
soliton-induced transparency has been coined
for this effect \cite{KozlovandKozlova2009}. To our knowledge, it has
previously been observed only in
the propagation of superradiance pulses in a neon 
plasma~\cite{Denisova98}.

In this paper, we show that soliton-induced transparency 
is readily seen in experiments using thermal vapor cells, in our case
a thin cell containing a rubidium vapor \cite{cell,james} addressed by a cw field 
resonant on the D1 transition and a pulsed field resonant on the D2 transition.
The numerical simulations described below reproduce the observed increase
in transmission and pulse shaping
indicating that this change signals the formation of a quasi-simulton.

\begin{figure}[b]
\centering
\includegraphics[width=0.48\textwidth]{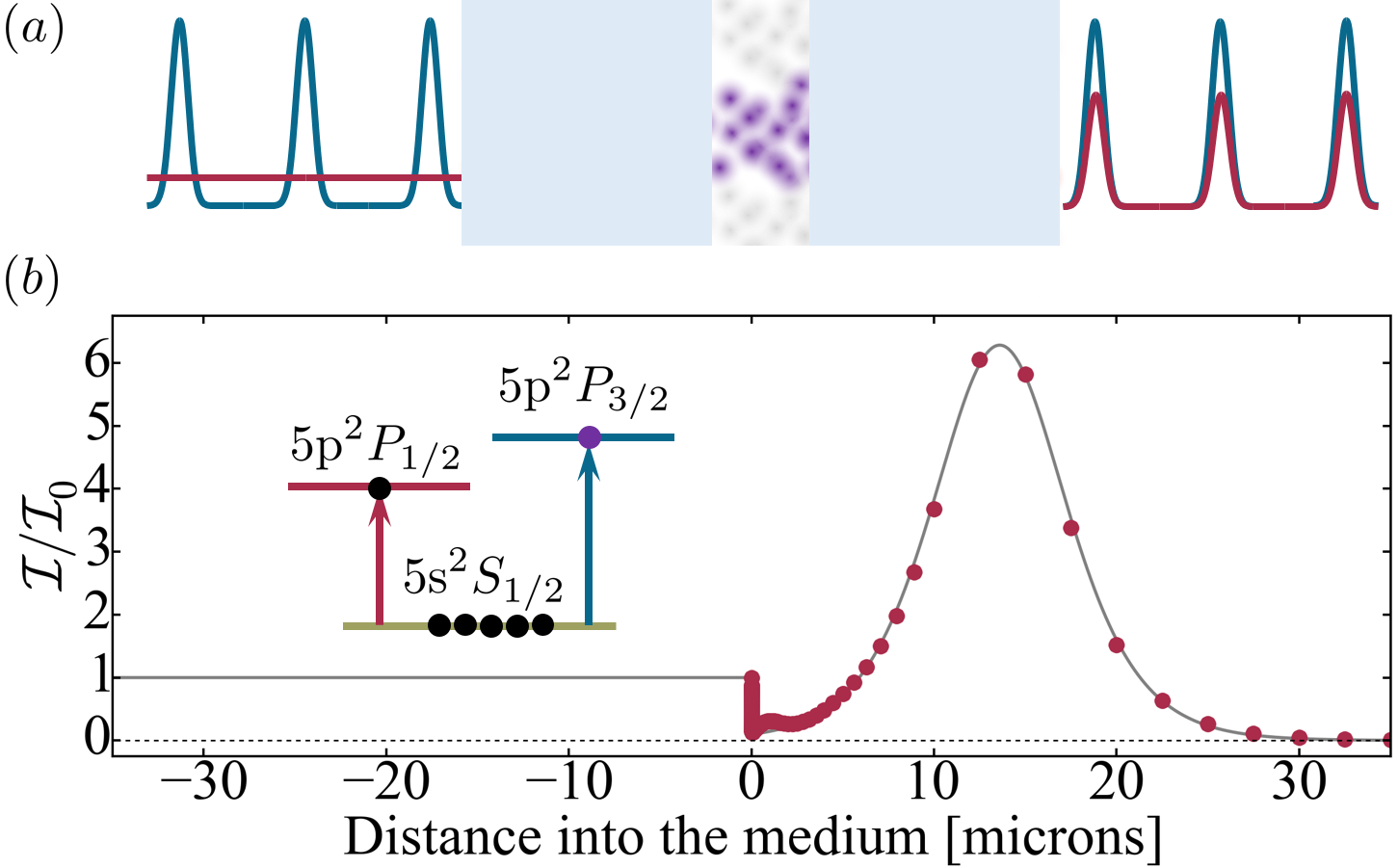}
\caption{(color online) (a) Schematic of the experiment (not to scale). A cw probe field (red)
resonant with the 5s$_{1/2}(F=3)$ -- 5p$_{1/2}(F=3)$ transition of $^{85}$Rb co-propagates with a pulsed coupling field (blue) resonant with the
5s$_{1/2}(F=3)$ -- 5p$_{3/2}(F=4)$ transition. These two beams are focused to a waist of order 10~$\mu$m inside a 2~$\mu$m vapor cell containing Rb. 
(b) Inset: The
level scheme indicating the probe and coupling fields. Main plot: Snapshot of the intensity of the probe field inside the medium, relative to the incident intensity, taken 0.05~ns after the coupling pulse reached its maximum at the front of the medium; the circles are calculated points and the solid line is a fit of these with a sech-squared pulse of width 8.6~$\mu$m (\textsc{fwhm}).}
\label{fig:schematic}
\end{figure}
\begin{figure}[t]
\centering
\includegraphics[width=0.48\textwidth]{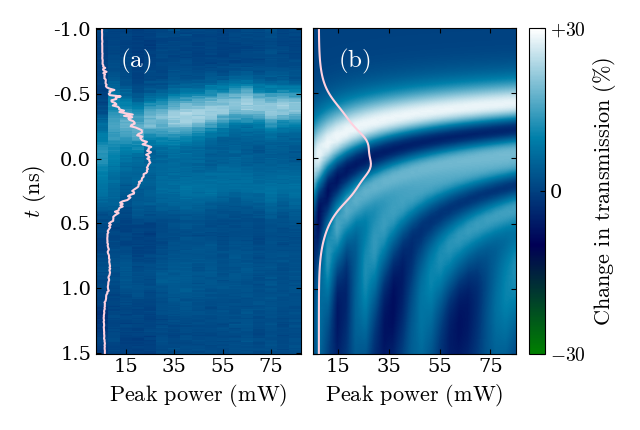}
\caption{\label{fig:data}(color online)
The change in the transmission at the probe frequency for a 2 $\mu$m-long
Rb vapor relative to
the average transmission at $t < -1$~ns,
vs.\ the peak power of the incident coupling pulse.
The pink curves represent the temporal intensity profile of the latter,
at the back of the medium, for an initial
peak power of 85 mW.
The
zero of the color scale corresponds to no change compared to the transmission
before the arrival of the coupling pulse.
The temperature is 200$^\circ$C.
(a): As measured.
(b): As predicted by the model described in the text.
}
\end{figure}

A sketch of the experiment is shown 
in Fig.~\ref{fig:schematic}(a).
A dense, 2-$\mu$m thick thermal vapor of rubidium atoms in their natural isotopic abundances
interacts with two co-propagating
monochromatic laser beams forming a V-type excitation scheme.
The probe and coupling beams are
linearly polarized in orthogonal directions. They are
resonant on, respectively,
the $5{\rm s}_{1/2}(F=3)$ -- $5{\rm p}_{1/2}(F=3)$ and
$5{\rm s}_{1/2}(F=3)$ -- $5{\rm p}_{3/2}(F=4)$
transitions of $^{85}$Rb. 
The coupling beam is
focused to a waist of $\sim\!20~\mu\mathrm{m}$ while the probe beam is
focused more tightly to a waist of $\sim\!10~\mu\mathrm{m}$, which minimizes
variation of the coupling intensity for the atoms in the probe
beam. The probe field applied to the cell is cw. The coupling field
is shaped to a short, nearly Gaussian pulse of a duration of typically 0.8~ns full width at half maximum (\textsc{fwhm}). Taking losses into account, we estimate
that at the front of the medium, and on axis,
the coupling field had an intensity of 
3.7~kW~cm$^{-2}$ at a measured peak power of 85~mW and
the probe field an intensity of
24~W~cm$^{-2}$. 
Following
propagation the two fields are separated by a polarizing beam splitter
and their transmission through the medium is monitored using a fast photodiode
\cite{Supplemental,Scully}. The temporal variation of the measured probe field
for various peak powers of the coupling pulse is shown in Fig.~\ref{fig:data}(a). The
main feature of these data is a strong
increase in transmission on the raising edge of the coupling pulse.

The other panel of this figure shows the results predicted by a model described below.
Before addressing these, 
we first consider a simplified model consisting
of a single ground
state (state 0) resonantly coupled to a first excited state by
a weak field (the probe field) and to a second excited state
by a strong pulse (the coupling field) [Fig.~\ref{fig:schematic}(b)]. 
We take
the coupling field at the front of the medium to be a Gaussian pulse
of 1~kW~cm$^{-2}$ peak intensity and
0.8~ns duration (\textsc{fwhm} in intensity), and the probe field
to have an initial constant intensity of 10~$\mu$W~cm$^{-2}$.
We set the excited states lifetimes and the transition wavelengths
and dipole moments to values corresponding to
the D1 and D2 lines of $^{85}$Rb, respectively.
The oscillator strengths of the two transitions thus differ by a factor of 2
here. We neglect Doppler broadening, self-broadening and
the hyperfine structure of the states for the time being.
We assume \textsc{1D} propagation \cite{transverse}, make the rotating wave and
slowly varying envelope approximations, and solve the resulting
Maxwell-Bloch equations numerically, taking the atoms to be
initially in the steady state driven by the probe field.

\begin{figure}[t]
\centering
\includegraphics[width=0.48\textwidth]{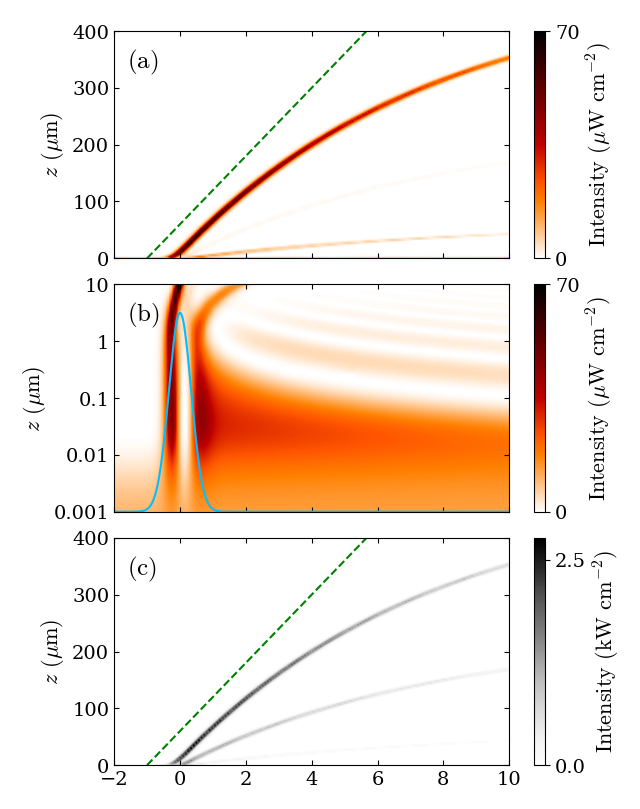}
\caption{(color online)
Formation of quasi-simultons in a 3-state model of an isotopically pure
$^{85}$Rb vapor,
neglecting Doppler broadening and self-broadening. The atomic density
corresponds to a temperature of 220$^\circ$C.
(a): The intensity of the probe field vs.\ time and propagation
distance within the medium. 
This field is
resonant on the D1 transition and has a constant intensity of
10 $\mu$W~cm$^{-2}$ at the entrance of the medium ($z=0$). 
The 0.8-ns coupling pulse is resonant on the D2 transition
and has a peak intensity of 1 kW~cm$^{-2}$ at $(t=0,z=0)$.
The dashed green line indicates the trajectory which a pulse propagating
at a constant speed of $2\times 10^{-4} \; c$  would trace in the figure.
(b): Enlargement of the small-$z$ region of panel (a). The
blue curve represents the temporal profile of 
the coupling pulse (the intensity scale is arbitrary).
(c): The same as (a) but for the coupling field.
}
\label{fig:3statelong}
\end{figure}
Fig.~\ref{fig:schematic}(b) and Fig.~\ref{fig:3statelong} show
how the probe and coupling fields vary both in space and in time
within this 3-state model.  Figs.~\ref{fig:3statelong}(a) and (b) refer to the 
probe field. As seen from these figures,
this field practically vanishes as soon as it enter the medium, prior to
the arrival of the coupling pulse
(the attenuation is extremely fast because the field is resonant with
the transition and Doppler broadening is neglected). 
However,
the arrival of that pulse triggers a more complicated dynamics. 
Microscopically, the onset of this dynamics
can be traced to a rapid increase 
of the $\rho_{12}$ coherence. This increase, in turns,
produces a large variation and a change
of sign of the $\rho_{01}$ coherence, leading to an amplification of the probe
field without a population inversion between the ground state and the first
excited state
\cite{Kozlov98,Denisova98}.
In
particular, the probe field develops into three successive pulses penetrating
far deeper into the medium than the initial cw field. 
As shown in
Fig.~\ref{fig:3statelong}(a),
these three pulses propagate soliton-like
over many tens of micrometers, each at a different speed
(the second of these three pulses is almost invisible in the figure).
The first one is the fastest, although its maximum speed is
only about
$2\times 10^{-4}\, c$. It is also the strongest, and at its peak
reaches an intensity larger than that of the incident cw field by
almost a factor of 7. 
Like the other two, it becomes weaker and
slower as it propagates. A snapshot of the spatial profile of
the probe field at a time when this first pulse has just formed is
shown in Fig.~\ref{fig:schematic}(b). The strong spatial localization of this pulse caused by
the slow light effect
is worth noting, as is its almost pure sech$^2$ profile.

The
0.8~ns coupling pulse also splits into three soliton-like pulses; the first
and second ones are well visible in
Fig.~\ref{fig:3statelong}(c), but not the third one. 
This pulse is strong enough
to break into three solitons even in the absence
of the probe field. The presence of the latter does not affect
the propagation of the coupling field significantly, but 
each of these three solitons co-propagates with a pulse 
at the probe frequency, thereby forming three quasi-simultons.

\begin{figure}[t]
\centering
\includegraphics[width=0.48\textwidth]{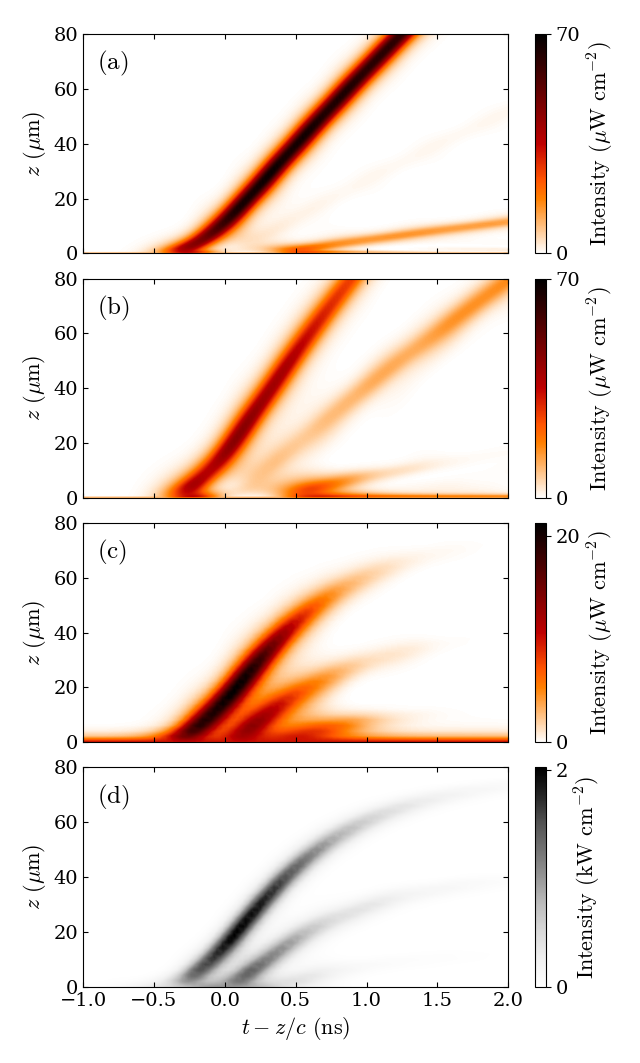}
\caption{(color online)
(a): The intensity of the probe field in
the 3-state model of Fig.~\ref{fig:3statelong}. This part of the figure
is an enlargement of the small $|t-z/c|$, small $z$ region of
Fig.~\ref{fig:3statelong}(b).
(b): The same as (a) but with
the full hyperfine structure of the 5s$_{1/2}$, 5p$_{1/2}$ and 
5p$_{3/2}$ states included in the calculation.
(c): The same as (b) but now also including Doppler broadening and
self-broadening.
(d): The same as (c) but for the coupling field.
}
\label{fig:all_08}
\end{figure} 
Differences in
transition dipole moments between the
magnetic substates coupled to each other by the two fields may compromise the formation of simultons
when the relevant levels have a hyperfine structure. However, this issue is not of major importance here
because the bandwidth
of the coupling pulse is sufficiently large compared to
the energy splitting between the relevant hyperfine states \cite{fewcycle}.
Fig.~\ref{fig:all_08} compares the
predictions of the 3-state model, in (a),
to the results obtained when
the complete hyperfine structure of the relevant levels is included
in the calculation, in (b).
While there are differences
between these two sets of results, it is clear that in the present case
the hyperfine structure does not prevent the formation of
quasi-simultons and their propagation over a considerable distance.  
However, the hyperfine structure 
of the 5s and 5p levels
is an important issue for longer coupling
pulses \cite{Supplemental,magneticfield}.

We now include not only the hyperfine structure but also Doppler broadening 
and self-broadening. The corresponding results are shown in panels (c) and
(d) of Fig.~\ref{fig:all_08}.
Quasi-simultons are still found in
these more complete calculations.
Although not as stable,
they still propagate over far longer
distances than would be the case for weak cw fields of the same wavelengths,
and the probe field is still 
amplified through its interaction with the
coupling pulse.

We used the same model as in Figs.~\ref{fig:all_08}(c) and (d) to obtain the results
displayed in Fig.~\ref{fig:data}(b), except that we assumed the incident probe field had
the same (much higher) intensity as in the experiment. We ignored the transverse Gaussian profiles of the laser beam, as factoring
these in would have led to excessively long computations. Interaction with the windows was taken into account
by broadening each state by 30~MHz \cite{JamesThesis}.
Comparing Fig.~\ref{fig:data}(b) to Fig.~\ref{fig:data}(a), we see that the model does
not predict the rapid damping of the dynamics which follows the initial increase in
transmission on the raising edge of the coupling pulse (the origin of this damping as yet unknown).
However, it reproduces this strong increase well.

\begin{figure}[t]
\centering
\includegraphics[width=0.48\textwidth]{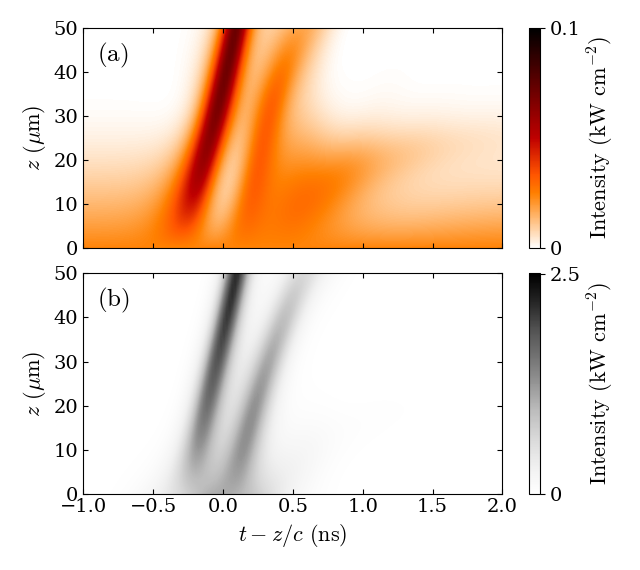}
\caption{\label{fig:exp80mum}(color online)
(a)
The calculated probe field for the conditions of Fig.~\ref{fig:data}, 
now for a much longer propagation distance of 50~$\mu$m. The peak power
of the coupling pulse is 35~mW. (b) The corresponding coupling field.
}
\end{figure}

We also ran this calculation for
a 50~$\mu$m-long cell, so as to see how this increase in the transmission
develops over longer propagation distances (Fig.~\ref{fig:exp80mum}).
Taking pulse re-shaping into account 
would be necessary for comparing to measurements
for a cell of that length; however, here we only aim at illustrating how
the \textsc{1D} dynamics would evolve beyond 2~$\mu$m if it was remaining
unperturbed. As seen from the figure, this enhancement would develop into a well defined pulse
co-propagating with the first of the solitons the coupling pulse
splits into. It can thus be identified with the formation
of a quasi-simulton.

\begin{figure}[t]
\centering
\includegraphics[width=0.48\textwidth]{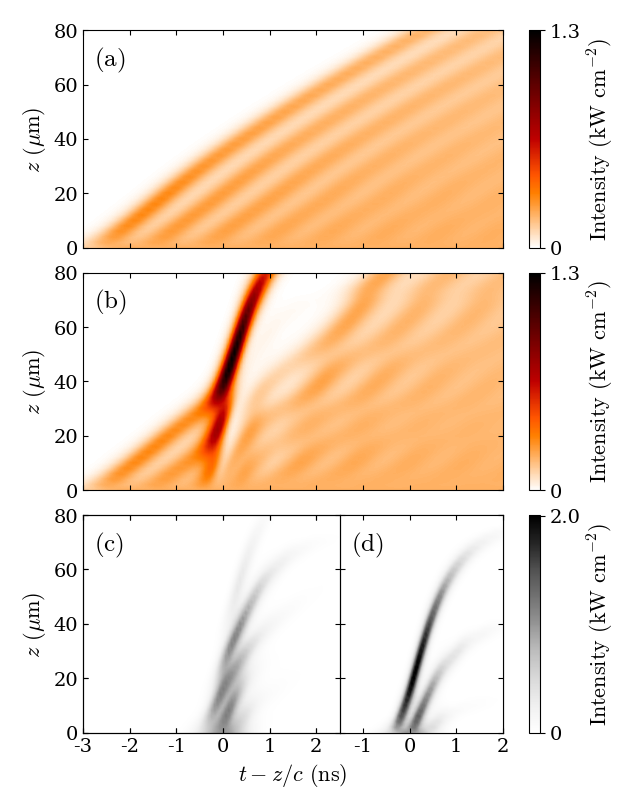}
\caption{\label{fig:composite}(color online)
(a) The probe field in the absence of a coupling pulse, assuming
that at $z = 0$ this field is a flat-top pulse smoothly turned on from 0 to
200~W~cm$^{-2}$ between $t=-5$ and $t=-2$~ns. 
As in Fig.~\ref{fig:all_08}(c), the model
includes Doppler and self-broadening as well as
the full hyperfine structure of the 5s and 5p energy levels. (b) The same
as (a) but here assuming an incident 0.8~ns coupling pulse
of 1~kW~cm$^{-2}$ peak intensity. (c): The coupling field corresponding to the probe field
shown in (b). (d): The same as (c) but assuming an incident intensity of only
10~$\mu$W~cm$^2$ at the probe frequency, instead of 200~W~cm$^{-2}$.
}
\end{figure}
It is interesting to explore this dynamics for still higher intensities 
of the probe field. To avoid strong optical pumping, we now assume that
the probe field at the entrance of the medium is
a flat-top pulse turned on over 3~ns. Taking the intensity after turn on to be
200~W~cm$^{-2}$ yields the results shown in panels (a), (b) and
(c) of Fig.~\ref{fig:composite}. Panel (a) shows the probe field
in the absence of a coupling pulse: at this high intensity, the
incident flat-top pulse splits into a periodic
train of soliton-like pulses upon entering the medium. Adding a strong coupling
pulse perturbs this dynamics
considerably, as shown in panel (b). In particular, we note the formation
and amplification of 
a strong pulse at the probe frequency associated with the pulse at the
coupling frequency. However, the latter [panel (c)]
is now strongly affected
by its interaction with the former,
contrary to what is found at lower intensities of the probe field [compare Fig.~\ref{fig:composite}(c) to Fig.~\ref{fig:composite}(d), which shows the coupling field found
when the probe intensity at $z=0$ is only 10~$\mu$W~cm$^{-2}$ after turn on].

In conclusion, we have demonstrated the amplification of a weak probe field
by a strong pulse --- the first step in the formation of quasi-simultons. Such simultons allow the propagation of weak localised fields through optical thick media. The satisfactory agreement between theory and experiment found in Fig.~\ref{fig:data} shows the applicability of our model, extended as necessary, to the exploration of what other two-color
propagation phenomena could be observed in thermal atomic vapors.
Future work will include investigating the single photon regime, the potential of Rydberg dressing \cite{rydberg} to induce strong photon-photon interaction, and the possibility of photon crystals.

\begin{acknowledgments}
CSA acknowledges financial support from EPSRC Grant Ref. Nos. EP/M014398/1, EP/P012000/1, EP/R002061/1, EP/R035482/1, and EP/S015973/1,
as well as DSTL and Durham University.
This work made use of the facilities of the Hamilton HPC Service of Durham University.
\end{acknowledgments}

\newpage

%
%
%
%

\begin{center}
{\large SUPPLEMENTAL MATERIAL}\\[3cm]
\end{center}

\section*{Method}
\subsection*{Experiment}
As is explained in the paper, the experimental setup involves two laser
beams (the probe beam and the coupling beam)
co-propagating through a thin cell containing
a Rb vapor.

The probe beam is first split by a polarizing 
beam-splitter cube (PBSC). Part of it is sent to a 
HighFinesse WS-7 wavemeter, resolution 0.1~pm. The wavelength 
is monitored through the latter
and controlled by a LABVIEW program locking the probe laser
to the 5s$_{1/2}(F=3)$ -- 5p$_{1/2}(F=3)$ transition of $^{85}$Rb.
The rest of the beam is sent to the cell.

The coupling beam is also first split
by a PBSC. The split beam is
used to lock the laser on the
5s$_{1/2}(F=3)$ -- 5p$_{3/2}(F=4)$ transition by
polarization spectroscopy \cite{Pearman}. The rest of the coupling beam
is sent to a Pockels cell immediately preceded and immediately followed
by crossed Glan-Taylor PBSCs with an extinction ratio exceeding
$10^{5}$. This very high extinction ratio
results in negligible beam power transmitted
through the Pockells cell when it is not activated. This cell is connected
to an electric pulse generator producing an approximately Gaussian pulse
with a \textsc{fwhm} of 0.8~ns. The temporal profile of the light pulse
so generated is recorded using a single-photon counting module. 

The two beams have orthogonal
linear polarizations, so that they can be separated with a PBSC
after the cell. In addition to this polarization filter
(extinction $\sim$ 200), we use a Semrock FF01-800/12 bandpass filter
passing 795~nm light with 95\% transmission and blocking 780~nm light
with a measured extinction of $3.5\times 10^3$. This setup ensures that
there is no detectable probe signal 
when the 780~nm coupling beam is on with
the probe beam off.

For detection, we use a fast photodiode with a 8~GHz bandwidth, forming
the input to a PicoScope 9221A 12~GHz bandwidth sampling
oscilloscope with an effective sampling rate of $\sim$ 400~GS~s$^{-1}$
(as the PicoScope is a sampling oscilloscope, not a real-time oscilloscope,
the data is an average over many pulse cycles). Systematic noise is
removed by recording signals with the probe laser off.

The cell has an inner length of 2~$\mu$m and is connected to a reservoir
of Rb in natural abundance. The assembly is heated to a temperature of 
200 $^\circ$C or higher so as to achieve a sufficiently high
atomic density.

The probe beam is focussed to a waist of $\sim$~10~$\mu$m whilst the
coupling beam is focussed less tightly to a waist of $\sim$~20~$\mu$m
to minimize any intensity variation of the coupling field
over the probe beam.
The same optics is used for focussing both beams.
The intensity of the beams
inside the cell is derived from spectroscopic measurements of Rabi
splitting in cw fields \cite{JamesThesis}.

\subsection{Theory}

We reduce the propagation problem to a form more easily amenable to numerical
calculation by making the rotating wave and slowly-varying envelope
approximations \cite{old}. We assume 
\textsc{1D} propagation in the $z$-direction and work in terms
of the total electric field vector 
\begin{equation}
{\bf E}(z,t) = \mbox{\boldmath ${\hat \epsilon}$}_{\rm p} E_{\rm p}(z,t)
+              \mbox{\boldmath ${\hat \epsilon}$}_{\rm c} E_{\rm c}(z,t),
\end{equation}
where the subscripts ${\rm p}$ and ${\rm c}$ refer to the probe 
and coupling fields and
$\mbox{\boldmath ${\hat \epsilon}$}_{{\rm p},{\rm c}}$ are unit
polarization vectors.
Writing the induced polarization field as 
\begin{equation}
{\bf P}(z,t) = \mbox{\boldmath ${\hat \epsilon}$}_{\rm p}P_{\rm p}(z,t)
+\mbox{\boldmath ${\hat \epsilon}$}_{\rm c}P_{\rm c}(z,t),
\end{equation}
we reduce the wave equation to the scalar form
\begin{equation}
{\partial^2 E_\alpha \over \partial z^2} - {1 \over c^2}\,
{\partial^2 E_\alpha \over \partial t^2} = \mu_0
{\partial^2 P_\alpha \over \partial t^2}, \quad \alpha = {\rm p},{\rm c}.
\end{equation}
We approximate each of these fields as the product of a
slowy-varying complex envelope and a carrier wave with
angular frequency $\omega_\alpha$ and wavenumber $k_\alpha$ ($\alpha = {\rm p},{\rm c}$):
\begin{align}
E_\alpha(z,t) &= {1 \over 2} {\cal E}_\alpha(z,t)\exp[i(k_\alpha z-\omega_\alpha t)] + \mbox{c.\ c.}, \\
P_\alpha(z,t) &= {1 \over 2} {\cal P}_\alpha(z,t)\exp[i(k_\alpha z-\omega_\alpha t)] + \mbox{c.\ c.}
\end{align}
Making the slowly-varying envelope approximation yields a 
first-order propagation equation for each field:
\begin{equation}
\left[
{\partial \; \over \partial z} + {1 \over c}\, 
{\partial \; \over \partial t}\right] {\cal E}_\alpha = 
{ik \over 2\epsilon_0}\,{\cal P}_\alpha, \quad \alpha = {\rm p},{\rm c}.
\end{equation}
The fields can thus be computed by integrating these equations subject to the relevant initial condition (at the entrance
of the medium), given the polarization fields ${\cal P}_{\rm p}(z,t)$ and ${\cal P}_{\rm c}(z,t)$.

We calculate the latter from the microscopic definition of the total polarization
as the expectation value of the 
dipole operator multiplied by the number density.
Without inhomogenous broadening,
\begin{equation}
{\bf P}(z,t) = {\cal N}\mbox{Tr}\,[{\bf d}\,\rho(z,t)],
\end{equation}
where ${\cal N}$ is the number of $^{85}$Rb
atoms per unit volume, ${\bf d}$ is the dipole operator and
$\rho(z,t)$ is the density operator representing the state of the atoms
driven by the field. When taking Doppler broadening into account we use
\begin{equation}
\label{eq:Doppler}
{\bf P}(z,t) = {\cal N} \int_{-\infty}^\infty f(v_z) \mbox{Tr}\,[{\bf d}\,\rho(z,t;
v_z)]\,{\rm d}v_z,
\end{equation}
where $v_z$ is the velocity in the $z$-direction, $\rho(z,t;v_z)$ is 
the density operator for atoms with that velocity, and
$f(v_z)$ is the Maxwell-Boltzmann probability distribution.
In either case,
we calculate the necessary coherences by solving the optical Bloch equations
within the rotating wave approximation.
We use either a 3-state model comprising only
a ground state and two excited states, as described in Fig.~1 of the paper,
or a model comprising all the hyperfine components of the
5s$_{1/2}$, 5p$_{1/2}$ and 5s$_{3/2}$ states (which brings the number
of coupled states from 3 to 46 \cite{46not48}).

We set
${\cal N} = 2.0\times 10^{15}$~cm$^{-3}$ in most of the calculations reported
in this publication, which is the number density corresponding to the
Rb vapor pressure for a temperature of 220 $^\circ$C assuming an isotopically
pure vapor.
The results shown in Figs.~2(b) and 5 of the paper were calculated
for ${\cal N} = 6.6\times 10^{14}$~cm$^{-3}$, the
number density of $^{85}$Rb atoms at 200 $^\circ$C for the
natural isotopic abundance, assuming no absorption by 
the $^{87}$Rb atoms present in the medium.

The matrix elements of the dipole operator are well known for
$^{85}$Rb \cite{Steck}:
\begin{align}
\langle 5&{\rm s}_{1/2}(F,m_F)\, |\, er_q\, |\, 5{\rm p}_{J'}(F',m_F')\rangle
= \nonumber \\
&\; \; (-1)^{F'}
\langle 5{\rm s}_{1/2}\,||\, e\,{\bf r}\,||\, 5{\rm p}_{J'}\rangle 
\langle F\,m_F | F'\,1\,m_F'\,q\rangle \nonumber \\
&\qquad \qquad \; \; \; \; \; \times 
\sqrt{2(2F'+1)} 
\left\{
\begin{matrix}
1/2 & J' & 1 \\ F' & F & 5/2
\end{matrix}
\right\},
\label{eq:F}
\end{align}
with 
$\langle 5{\rm s}_{1/2}\,||\, e\,{\bf r}\,||\, 5{\rm p}_{1/2}\rangle 
= 2.54\times 10^{-29}$~C~m and
$\langle 5{\rm s}_{1/2}\,||\, e\,{\bf r}\,||\, 5{\rm p}_{3/2}\rangle 
= 3.58\times 10^{-29}$~C~m.
We entirely neglect 
the hyperfine structure of the 5s and 5p levels
in the 3-state model of Figs.~3 and 4(a) of the paper; instead,
we assume $z$-polarization and $m_J = m_J' = 1/2$, and use
\begin{align}
\langle 5{\rm s}_{1/2},  m_J = 1/2\, |\, er_0\, & |\, 5{\rm p}_{J'}, m_J'=1/2\rangle
= \nonumber \\
&-\langle 5{\rm s}_{1/2}\,||\, e\,{\bf r}\,||\, 5{\rm p}_{J'}\rangle / \sqrt{3}.
\label{eq:J}
\end{align}
The natural frequency widths of the 5p$_{1/2}$ and
5p$_{3/2}$ states are, respectively,
5.75 and 6.07~MHz. 
We treat
self-broadening (collisional broadening)
as an additional dephasing term, using the
same decay widths as in Ref.~\cite{Leepaper} and the same branching 
ratios between states as for spontaneous decay.

What state the atoms are in prior to
the arrival of the pulse at the coupling frequency matters considerably.
In the 3-state model, each atom is initially in the
$m=1/2$ ground state, and we let this state evolve into a 
stationary coherent superposition of the 5s$_{1/2}$ and
5p$_{1/2}$ states under the effect of the probe field alone before
applying the coupling pulse \cite{note}. For the very weak
probe field considered in Fig.~3 of the paper,
it makes practically no difference
whether this stationary state is taken to be the exact solution 
of the optical Bloch equations in the long time limit
or the solution obtained
within the weak field approximation. (The latter is that 
obtained by finding the stationary
solution of the optical Bloch equations subject to the constraint
that the excited states have zero population.) The results shown in Fig.~3,
which
were obtained within the weak field approximation, cannot be distinguished
from the results obtained without 
this approximation on the scale of the figure.

However, making 
or not making the weak field approximation yields
markedly different stationary states once the hyperfine structure of the
levels is built in the model.
We assume that the
atoms are at first uniformly distributed between the twelve
5s$_{1/2}(F=2)$ and 5s$_{1/2}(F=3)$ states.
Making the weak field approximation yields a
density matrix with populations of 1/12 for each of
the 5s$_{1/2}(F=3)$ states,
whereas not making it reduces
these populations to much lower values
through optical pumping to the 5s$_{1/2}(F=2)$
states. The medium is then almost transparent to
both fields, which changes the dynamics dramatically
(Fig.~S1).
\begin{figure}[t]
\centering
\includegraphics[width=0.48\textwidth]{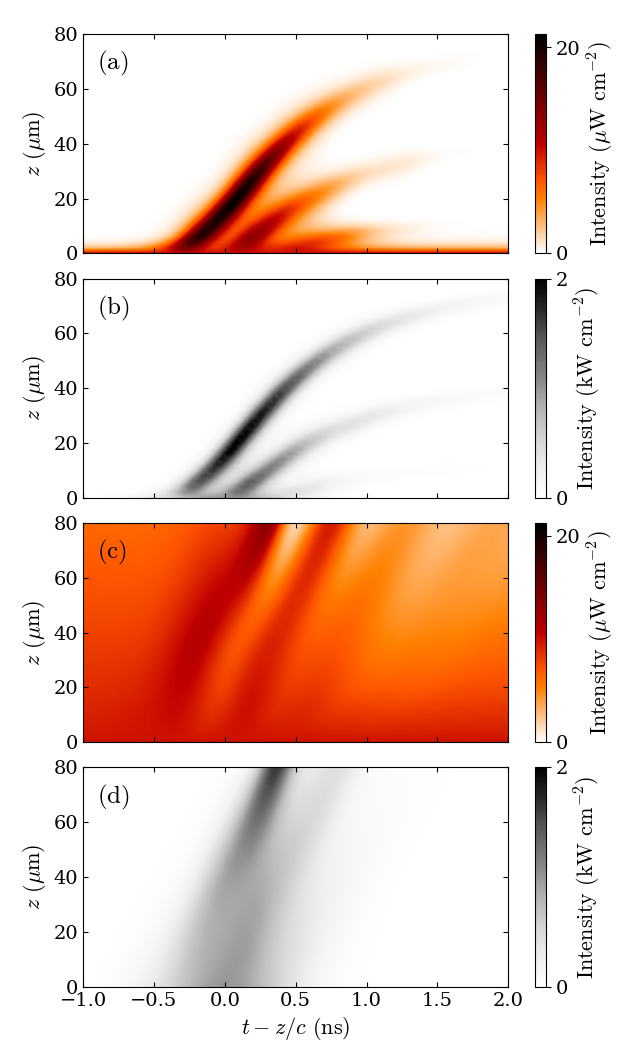}
\caption{
Predictions of the 46-state model including Doppler broadening and
self-broadening for two different states of the medium prior to the
arrival of the coupling pulse.
(a) and (b): the medium is initially in the steady state obtained
in the weak field approximation.
(c) and (d): the medium is initially in the steady state predicted
by the full optical Bloch equations.
(a) and (c): the intensity of the probe field.
(b) and (d): the intensity of the coupling field.
}
\end{figure}

We generally use the weak field approximation
when comparing to the 3-state model, so as to compare like to like.
Doing so would not be appropriate when comparing to the data, though, given the strength of the probe field used in the experiment. As the atoms evolve rapidly into a state close if not practically identical to a steady, optically pumped state when they cross the beams, we take the state of the medium prior to the arrival of the coupling pulse to be
the exact stationary solution of the optical Bloch equations for the probe field alone. Broadening each state by 30~MHz on account of the interaction with the windows of the nanocell \cite{JamesThesis} reduces optical pumping, which leads to a larger absorption as would otherwise be the case --- compare Figs.~S1(c) and (d) to Figs.~5(a) and (b) of the paper.     

\begin{figure}[t]
\centering
\includegraphics[width=0.48\textwidth]{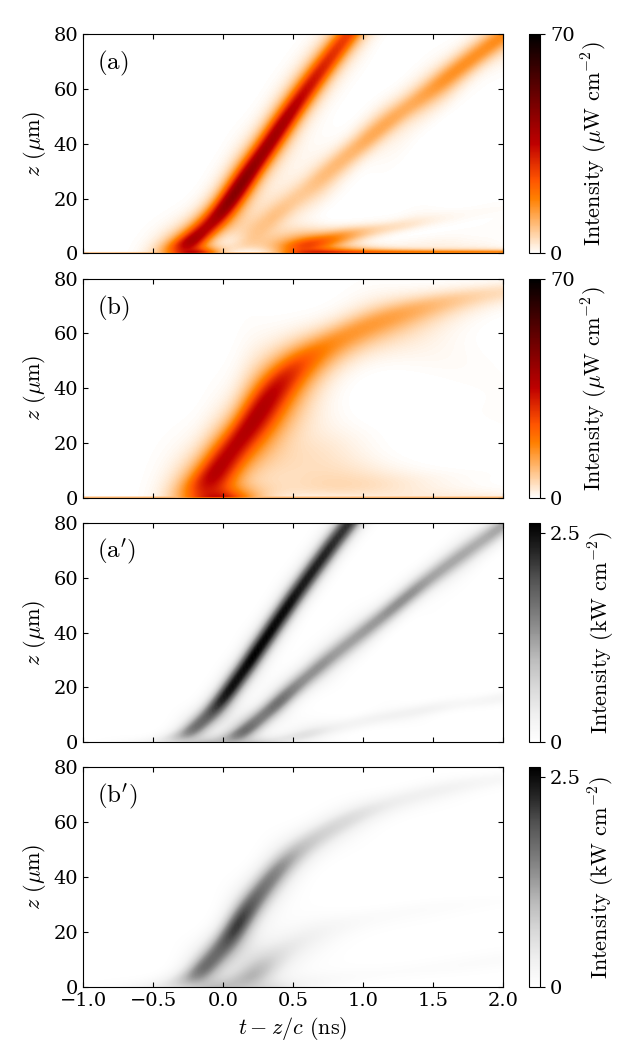}
\caption{(a) and (a'):
the fields obtained within the 46-state model of Fig.~4(b)
of the paper. (b) and (b'):
the fields obtained when the non-resonant hyperfine
levels are ignored. Top two panels: the intensity of the
probe field. Bottom two panels: the intensity of the coupling field.
}
\end{figure}

\subsection*{Computations}

Like previous investigators \cite{old},
we worked in terms of the distance $z' = z$ and 
the retarded time $\tau =  t - z/c$, in view of the fact that
\begin{equation}
{\partial \; \over \partial z} + {1 \over c}\,
{\partial \; \over \partial t} \rightarrow {\partial \; \over \partial z'}
\end{equation}
under this change of variables. We divided the relevant ranges of
values of $z'$ and $\tau$ into a sufficient number of steps --- e.g., 16 steps
in $z'$ in the
case of Fig.~2(b) of the paper, but as many as 32,000 in the case of
Figs.~3(b) and 3(c). We calculated the fields by integrating
the Maxwell-Bloch equations
over $z'$, starting at $z' = 0$ (the front
of the medium, where the fields have given values):
knowing the fields and
the polarization of the medium at $z'$ (the polarization
being calculated as described below), we calculated the fields at 
$z' + \delta z'$.
We normally
used a non-adaptative predictor-corrector approach for this, in which
a third-order Adams-Bashforth step was followed by
a fourth-order Adams-Moulton step, the fourth-order Runge-Kutta rule
being used to start the integration. In some cases, where convergence tests
indicated this to be more efficient, we
used the second-order Adams-Bashforth rule instead
(started with the mid-point formula), without an Adams-Moulton step.

At each step in $z'$ we obtained the polarization of the medium by
solving the optical Bloch equations. To this effect we
used either the fourth-order
Runge-Kutta formula or
Butcher's fifth-order Runge-Kutta formula \cite{Butcher}, whichever was 
the most efficient for the case at hand. The integral over $v_z$ appearing
in Eq.~(\ref{eq:Doppler})
was calculated
using a 24-point Clenshaw-Curtis rule, which was sufficiently accurate
at the temperatures considered.

\section{Role of the hyperfine structure}

Fig.~S2 illustrates the importance of taking into account the whole of
the hyperfine structure of the levels, not just that of the resonantly
coupled
5s$_{1/2}(F=3)$, 5p$_{1/2}(F=3)$ and 5p$_{3/2}(F=4)$ states.
Panels (a) and (a') of this figure show the intensity distribution
of the fields inside the medium as calculated with the entire hyperfine
structure included (46 states in total), whereas panels (b) and (b') show
the results obtained when only the Zeeman substates of the
5s$_{1/2}(F=3)$, 5p$_{1/2}(F=3)$ and 5p$_{3/2}(F=4)$ states are included. 
It is clear from Fig.~S2 that 
neglecting the 5s$_{1/2}(F=2)$, 5p$_{1/2}(F=2)$ and 5p$_{3/2}(F=1,2,3)$ states
reduces the stability of the quasi-simultons or even prevents their
formation. This difference originates from the fact that different
pairs of states differing by the values of $m_F$ and $m_F'$ have different
transition dipole moments, which compromises the coherent propagation of
the two fields through the medium. However, the bandwidth of the coupling
pulse ($\sim$ 0.4~GHz for a duration of 0.8~ns) is larger than
the hyperfine splitting of the 5p$_{3/2}$ level (0.21~GHz) and comparable
to that of the 5p$_{1/2}$ level (0.36~GHz). The atoms thus
tend to couple to the field
as if these levels had no hyperfine structure. If the hyperfine splitting
was
really zero, and ignoring the 5s$_{1/2}(F=2)$ states (which are far
off resonance),
the system would be equivalent to
the 3-state system discussed in the paper
and there would be only a single transition dipole moment
for each of the two frequencies. Mathematically, the equivalence
arises from 
the following sum rule,
\vfill\eject
\begin{align}
\sum_{F'}
|\langle 5{\rm s}_{1/2}(F,m_F)&\, |\, er_0\, |\, 5{\rm p}_{J'}(F',m_F)\rangle|^2
= \nonumber \\
&|\langle 5{\rm s}_{1/2}\,||\, e\,{\bf r}\,||\, 5{\rm p}_{J'}\rangle|^2 /3,
\end{align}
which applies for both $F=3$ and $F=2$
and relates the dipole moments
of Eq.~(\ref{eq:F}) to those of Eq.~(\ref{eq:J}).

The equivalence with the 3-state model
is not exact, though, both because the hyperfine splitting
of the 5p states is not completely negligible compared to the
bandwidths of the pulse and because 
only 7/12 of the population is initially in the 5s$_{1/2}(F=3)$ state in
the 46-state model. By contrast, 100\% of the population is initially
in the ground state in the 3-state model.
The density of atoms
resonantly coupled to the 5p states by the two fields being smaller
in the 46-state model,
the quasi-simultons
tend to propagate faster in that model than in the 3-state
model --- e.g., compare Fig.~4(b) of the paper to Fig.~4(a).

By the same token, decreasing the bandwidth of the coupling pulse by
increasing its duration also
reduces the stability of the quasi-simultons. This is illustrated by Fig.~S3.
The models, field strengths and coupling pulse duration are the same in
panels (a), (a'), (d) and (d') as
in Figs.~4(a) and 4(b) of the paper.
Compared to the first column of the figure,
the coupling pulse is 2.5 times
longer in the second column and 5 times longer in the third
column. The peak intensity of this pulse is reduced correspondingly
so as to keep its initial area the same (in all three cases,
this pulse splits into three solitons in the
3-state model). As seen from the figure, the
agreement between the 3-state and 46-state results degrades significantly
as the hyperfine splitting of the states starts exceeding the
bandwidth of the coupling pulse.
For a pulse duration of 4~ns, the probe and coupling fields do no
longer co-propagate in the form of clearly defined quasi-simultons.
\begin{turnpage}
\begin{figure}[t]
\centering
\includegraphics[width=1.18\textwidth]{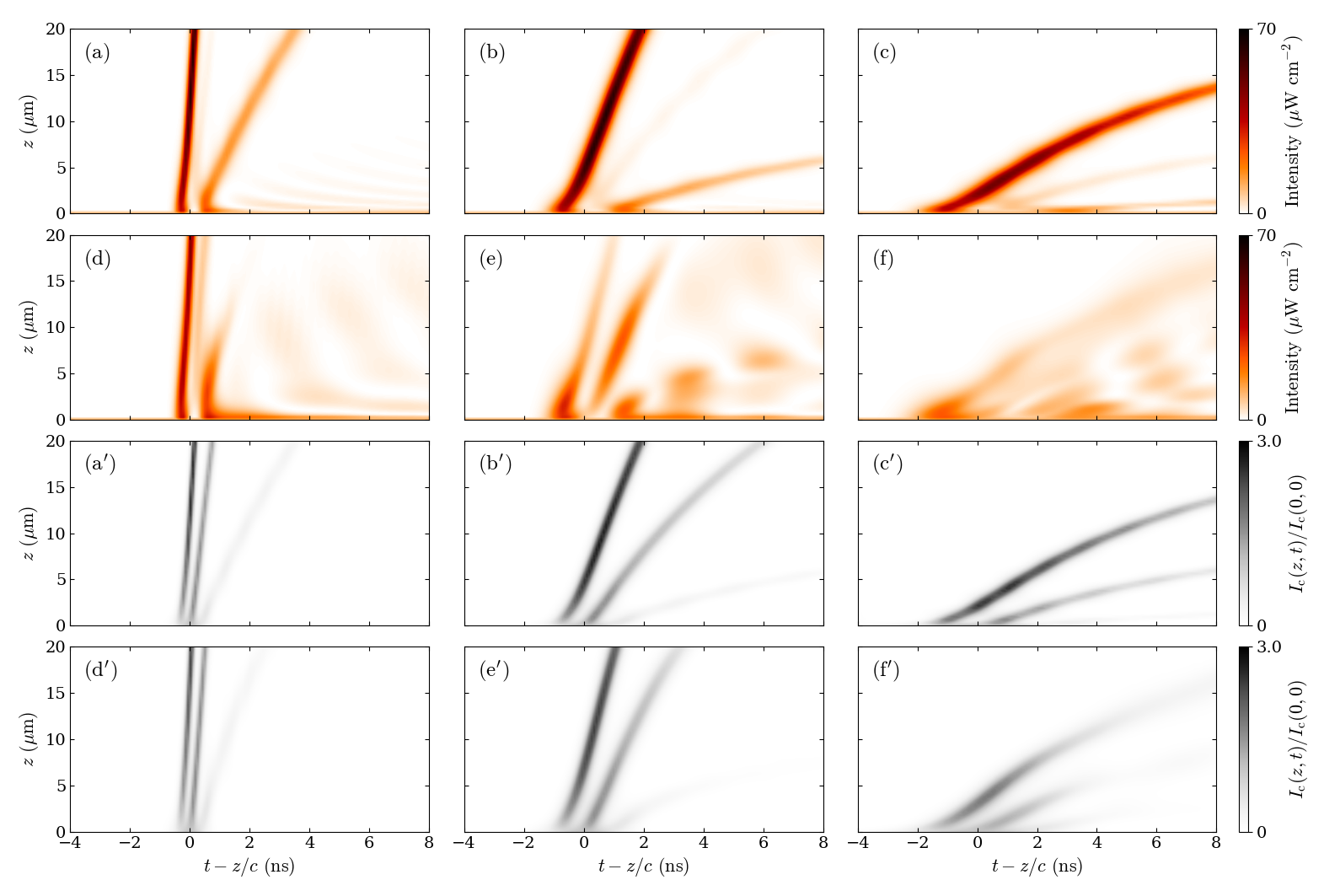}
\caption{Comparison between the 3-state model
of Figs.~3 and 4(a) of the paper and the 46-state model
of Fig.~4(b) for different lengths of the coupling pulse.
Top two rows:
the intensity of the probe field.
Bottom two rows: the intensity of the
coupling field. 
(a), (b) and (c), (a'), (b') and (c'):
the 3-state model.
(d), (e) and (f), (d'), (e') and (f'): the 46-state model.
(a), (d), (a') and (d'): $\tau_{\rm c} = 0.8$~ns and
$I_{\rm c}(z=0,t=0) = 1$~kW~cm$^{-2}$.
(b), (e), (b') and (e'): $\tau_{\rm c} = 2$~ns and
$I_{\rm c}(z=0,t=0) = 0.16$~kW~cm$^{-2}$.
(c), (f), (c') and (f'): $\tau_{\rm c} = 4$~ns and
$I_{\rm c}(z=0,t=0) = 0.04$~kW~cm$^{-2}$.
}
\end{figure}
\end{turnpage}

\


\begin{thebibliography}{9}
\bibitem{46not48}
There are 48 hyperfine 5s and 5p states in total. However,
taking the quantization axis parallel to the polarization direction of 
the coupling field, the 5p$_{3/2}(F=4,m_F = \pm 4)$ states are not coupled
to any of the other 46 states.
\bibitem{note}
Adding the $m=-1/2$ states to this model and assuming that
half the atoms
are initially in the $m=1/2$ ground state and half in the
$m=-1/2$ ground state would lead to exactly the same results.
\end{thebibliography}
\end{document}